\documentclass[fleqn]{article}
\usepackage{longtable}
\usepackage{graphicx}

\begin {document}
\begin{flushleft}
{\LARGE
{\bf Comment on ``Collision strength and effective collision strength for Ba~XLVIII" by Mohan et al.  [Can. J. Phys.   95 (2017) 173]}
}\\

\vspace{1.5 cm}

{\bf {Kanti  M  ~Aggarwal}}\\ 

\vspace*{1.0cm}

Astrophysics Research Centre, School of Mathematics and Physics, Queen's University Belfast, \\Belfast BT7 1NN, Northern Ireland, UK\\ 
\vspace*{0.5 cm} 

e-mail: K.Aggarwal@qub.ac.uk \\

\vspace*{0.20cm}

Received: 26 October 2017; Accepted: 23 May 2018

\vspace*{1.0 cm}

\vspace*{1.0 cm}
\vspace*{1.0 cm}

\hrule

\vspace{0.5 cm}

\end{flushleft}

\clearpage


\begin{abstract}

In a recent paper, Mohan et al. [Can. J. Phys.  {\bf 95} (2017) 173] have reported results for  collision strengths  ($\Omega$) and effective collision strengths ($\Upsilon$) for transitions from the ground to  higher 51 excited levels of F-like Ba~XLVIII. For the calculations of $\Omega$,  the Dirac atomic $R$-matrix code (DARC) and the flexible atomic code (FAC)  have been adopted, in order to facilitate a direct comparison. However, for the subsequent calculations of $\Upsilon$, DARC alone has been employed. In this comment, we demonstrate that while their limited results for $\Omega$ are comparatively reliable, for $\Upsilon$ are not, particularly for the allowed transitions and at lower temperatures.   Apart from the non expected behaviour, their $\Upsilon$ values are overestimated for several transitions,  by about a factor of two.

\end{abstract}

\clearpage

\section{Introduction}

In a recent paper, Mohan et al. \cite{mm1} have reported results for  collision strengths  ($\Omega$) and effective collision strengths ($\Upsilon$) for transitions from the 2s$^2$2p$^5$~$^2$P$^o_{3/2}$ ground to 51 excited levels  of F-like Ba~XLVIII, which  belong to the 2s$^2$2p$^5$, 2s2p$^6$, and 2s$^2$2p$^4$3$\ell$ configurations. For the calculations of $\Omega$, they have adopted two independent codes, namely the Dirac atomic $R$-matrix code (DARC) and the flexible atomic code (FAC). These codes are freely available on the websites \\{\tt http://amdpp.phys.strath.ac.uk/UK\_APAP/codes.html} and {\tt https://www-amdis.iaea.org/FAC/}, respectively. By making comparisons between the two calculations for $\Omega$, although for only limited ($<$4\%) transitions, they have concluded a good agreement, for most transitions. However, for the subsequent calculations of $\Upsilon$, DARC alone has been employed. In the absence of any other existing similar results, no direct comparisons were possible, but they speculated their data to be `reliable, authentic, and accurate'.   However, we  find exactly the opposite of this, because their $\Upsilon$ results are neither correct in behaviour nor in magnitude. 

\section {Collision strengths and effective collision strengths}

For the construction of  wavefunctions, Mohan et al. \cite{mm1} adopted the general-purpose relativistic atomic structure package (GRASP). Several versions of this code are currently in use by many workers, but the one adopted by them is the original one, i.e. GRASP0, but considerably modified by P.H.~Norrington and I.P.~Grant. This is available at the same website as DARC, and can be directly linked to the latter. They included a large CI (configuration interaction) among 431 levels of 29 configurations, namely  2s$^2$2p$^5$, 2s2p$^6$, 2s2p$^5$3$\ell$,  2p$^6$3$\ell$, 2s$^2$2p$^4$3$\ell$, 2s$^2$2p$^4$4$\ell$, 2s2p$^5$4$\ell$, 2s$^2$2p$^4$5$\ell$, and 2s2p$^5$5$\ell$. Energies for these levels and oscillator strengths (f-values) for transitions among these have already been reported by them in a separate paper \cite{mm2}.  Since inclusion of such a large number of levels in a scattering calculation requires significantly large computational resources, they restricted their work  to the lowest 52 levels, which belong to the 2s$^2$2p$^5$, 2s2p$^6$, and 2s$^2$2p$^4$3$\ell$  configurations. In fact, these configurations generate 60 levels in total, but they have preferred to ignore the remaining 8. Nevertheless, CI for F-like ions is not very important as may be noted from our work on 17  ions with 37 $\le$ Z $\le$ 53 \cite{kma1}. Therefore, we have performed  simple calculations among these 60 levels, which will facilitate a direct comparison with their results, and hence some assessment of the accuracy and reliability of their data.  

For the calculations of $\Omega$, Mohan et al. \cite{mm1}  adopted the DARC code, included a wide range of partial waves with angular momentum $J \le$ 41, considered a wide range of energy (up to 1000~Ryd), and resolved resonances to determine $\Upsilon$ at temperatures below 10$^6$~K. For our calculations, the FAC is employed which is based on distorted-wave (DW) approximation, and generally provides comparable results of $\Omega$ with $R$-matrix, as may also be noted from Table~2 of \cite{mm1}. Similarly, the energies obtained with this code are comparable with those of Mohan et al., and there are no appreciable discrepancies, in magnitude or orderings, for any level -- see also Table~1 of \cite{mm1}. For this reason we are not listing our energy levels for Ba~XLVIII. However, in our subsequent calculations of $\Upsilon$, resonances are not included, which mainly affect the forbidden transitions, not the allowed  ones. For this reason we will focus our comparisons on some allowed transitions alone.

\begin{table}
\caption{Comparison of oscillator strengths (f-values) for some transitions of Ba~XLVIII from ground to higher excited levels.} 
\begin{tabular}{rrcrrr}  \hline
 I  &  J & Transition  & GRASP &  FAC  \\
 \hline
    1  &   22  &  2s$^2$2p$^5$~$^2$P$^o_{3/2}$ -- 2s$^2$2p$^4$($^3$P$_2$)3d~$^2$P$_{1/2}$ &  0.120  &  0.119   \\
    1  &   23  &  2s$^2$2p$^5$~$^2$P$^o_{3/2}$ -- 2s$^2$2p$^4$($^3$P$_2$)3d~$^2$D$_{5/2}$ &  0.624  &  0.636   \\
    1  &   24  &  2s$^2$2p$^5$~$^2$P$^o_{3/2}$ -- 2s$^2$2p$^4$($^3$P$_2$)3d~$^2$P$_{3/2}$ &  0.463  &  0.437   \\
    1  &   25  &  2s$^2$2p$^5$~$^2$P$^o_{3/2}$ -- 2s$^2$2p$^4$($^1$S$_0$)3d~$^2$D$_{5/2}$ &  0.464  &  0.455   \\
\hline	
\end{tabular}

\begin{flushleft}
{\small
GRASP: calculations of Mohan et al. \cite{mm1},\cite{mm2} with GRASP \\
FAC: present calculations with FAC  \\
}
\end{flushleft}
\end{table}

\begin{figure*}
\includegraphics[angle=-90,width=0.9\textwidth]{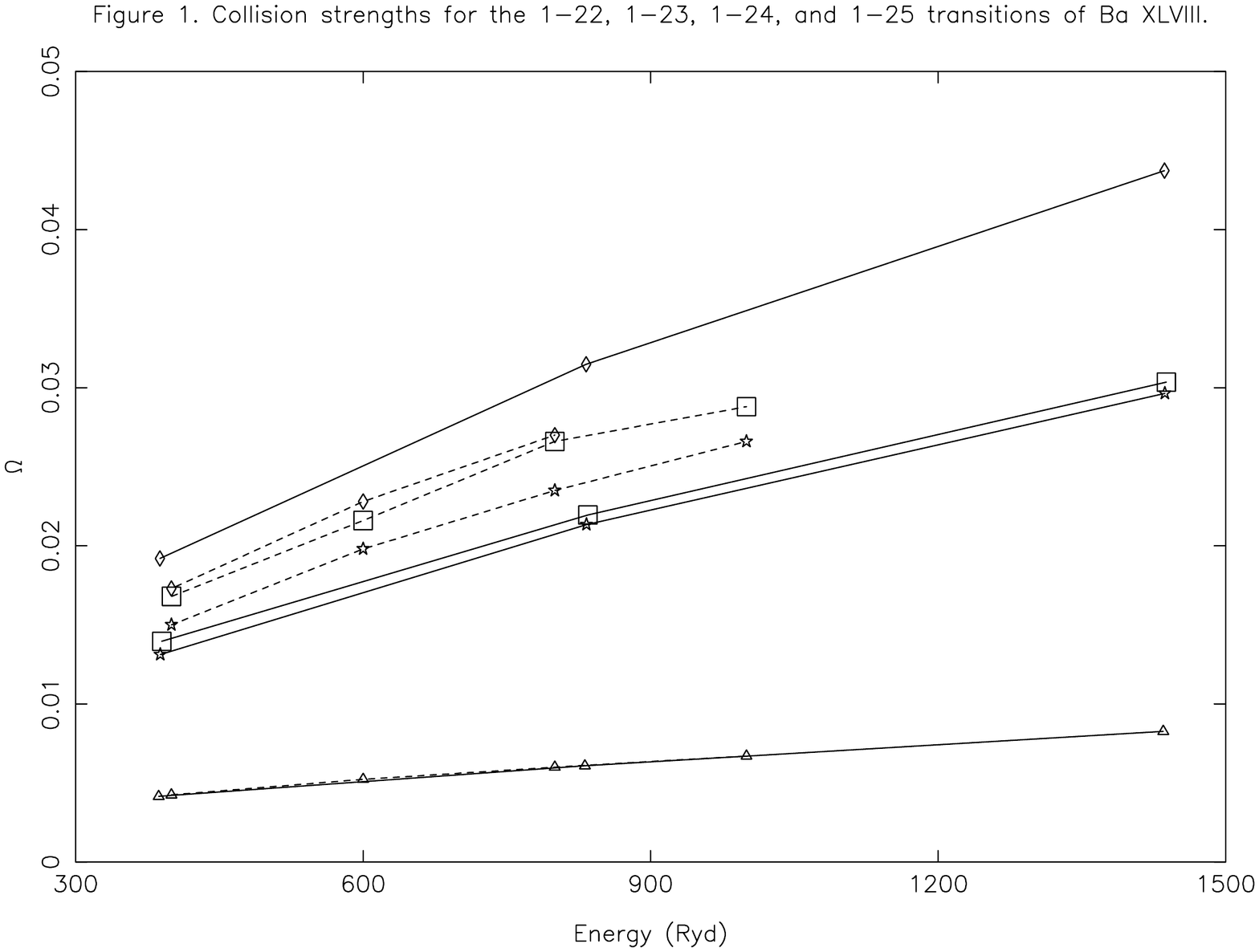}
 \vspace{-1.5cm}
 \caption{Comparison of DARC and FAC values of $\Omega$ for  the 1--22 (2s$^2$2p$^5$~$^2$P$^o_{3/2}$ -- 2s$^2$2p$^4$($^3$P$_2$)3d~$^2$P$_{1/2}$: triangles), 1--23 (2s$^2$2p$^5$~$^2$P$^o_{3/2}$ -- 2s$^2$2p$^4$($^3$P$_2$)3d~$^2$D$_{5/2}$: diamonds), 1--24 (2s$^2$2p$^5$~$^2$P$^o_{3/2}$ -- 2s$^2$2p$^4$($^3$P$_2$)3d~$^2$P$_{3/2}$: stars), and 1--25 (2s$^2$2p$^5$~$^2$P$^o_{3/2}$ -- 2s$^2$2p$^4$($^1$S$_0$)3d~$^2$D$_{5/2}$: squares) transitions of Ba~XLVIII. Continuous curves: present results with FAC, broken curves: earlier results of Mohan et al. \cite{mm1} with DARC.}
 \end{figure*}

In Table~1 we compare f-values between our work and that of Mohan et al. \cite{mm1} for four transitions, namely 1--22/23/24/25, i.e. 2s$^2$2p$^5$~$^2$P$^o_{3/2}$ -- (2s$^2$2p$^4$)[$^3$P$_2$]3d~$^2$P$^o_{1/2}$, [$^3$P$_2$]3d~$^2$D$^o_{5/2}$, [$^3$P$_2$]3d~$^2$P$^o_{3/2}$, and [$^1$S$_0$]3d~$^2$D$^o_{5/2}$ -- see Table~1 of \cite{mm1} for the definition of all levels. These transitions are allowed, comparatively stronger, and have no discrepancies in their f-values. Therefore, the corresponding results for $\Omega$ are also expected to be comparable (within a few percents), which indeed is the case as shown in Fig. 1. In Fig.~2 we make similar comparisons for $\Upsilon$, but the differences are striking between the two independent calculations, particularly at temperatures below 10$^5$~K. Since resonances (if any) do not make a significant contribution to the determination of $\Upsilon$, the reported results of Mohan et al. are not only abnormal in behaviour towards the lower end of the temperature range, but are also overestimated, by up to a factor of two. 
 
 \begin{figure*}
\includegraphics[angle=-90,width=0.9\textwidth]{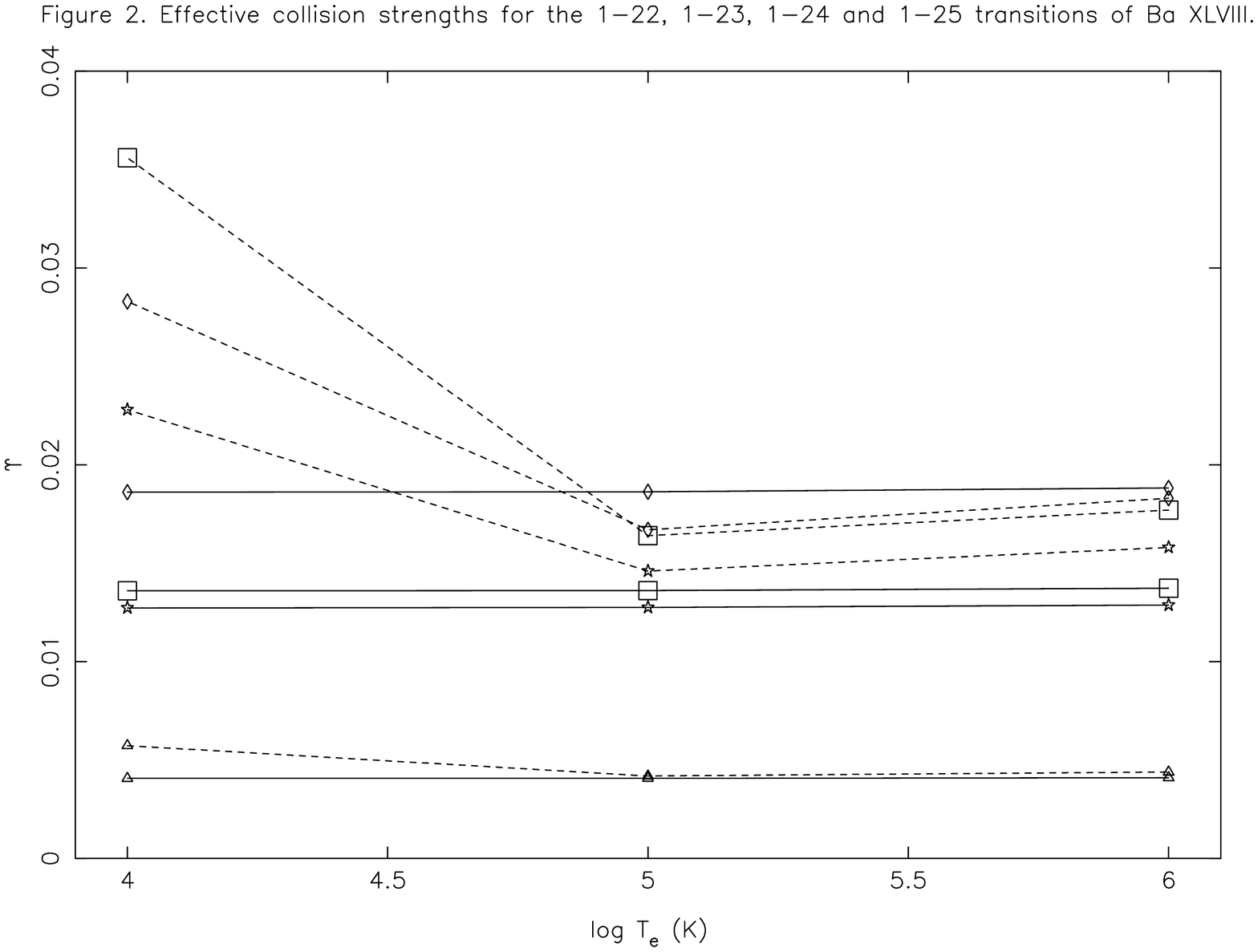}
 \vspace{-1.5cm}
 \caption{Comparison of DARC and FAC  values of $\Upsilon$ for the 1--22 (2s$^2$2p$^5$~$^2$P$^o_{3/2}$ -- 2s$^2$2p$^4$($^3$P$_2$)3d~$^2$P$_{1/2}$: triangles), 1--23 (2s$^2$2p$^5$~$^2$P$^o_{3/2}$ -- 2s$^2$2p$^4$($^3$P$_2$)3d~$^2$D$_{5/2}$: diamonds), 1--24 (2s$^2$2p$^5$~$^2$P$^o_{3/2}$ -- 2s$^2$2p$^4$($^3$P$_2$)3d~$^2$P$_{3/2}$: stars), and 1--25 (2s$^2$2p$^5$~$^2$P$^o_{3/2}$ -- 2s$^2$2p$^4$($^1$S$_0$)3d~$^2$D$_{5/2}$: squares) transitions of Ba~XLVIII. Continuous curves: present results with FAC, broken curves: earlier results of Mohan et al. \cite{mm1} with DARC.}
 \end{figure*}
 
 Another example for the incorrect $\Upsilon$ results of Mohan et al. \cite{mm1} is the 1--3 transition, i.e. 2s$^2$2p$^5$~$^2$P$^o_{3/2}$ -- 2s2p$^6$~$^2$S$_{1/2}$, for which they have shown $\Omega$ in their Fig.~2. It is clear that $\Omega$ values rise gradually over a wide energy range of about 100~Ryd, without any resonances. Therefore, corresponding $\Upsilon$ results should also either rise gradually, or remain nearly constant, over the small temperature range below 10$^6$~K, equivalent to $\sim$6.3~Ryd. This indeed is the case in our calculations as $\Upsilon$ is 0.053, whereas the corresponding results of Mohan et al. are 0.0767, 0.0516, and 0.0587 at the respective temperatures of 10$^4$, 10$^5$, and 10$^6$~K. The same conclusion of invariable $\Upsilon$, for this transition at temperatures below 10$^6$~K, was noted earlier  for another F-like ion, namely  Kr~XXVIII \cite{kr28}. Therefore, for the 1--3 transition the $\Upsilon$ value of Mohan et al. is clearly overestimated by 50\% at T$_e$ = 10$^4$~K. 
 
 The reason for the anomalous behaviour of $\Upsilon$ results by Mohan et al. \cite{mm1}, particularly at lower temperatures,  is not difficult to diagnose. This is primarily due to their choice of  energy mesh ($\delta$E) for the resolution of resonances and the determination of $\Upsilon$. Their $\delta$E is 0.065~Ryd  equivalent to $\sim$10~260~K, i.e. greater than the lowest T$_e$ (10~000~K)  they considered. For an accurate performance of the integral in Eq.~13 of \cite{mm1}, this  energy mesh is (very) very coarse, and hence has greatly affected the $\Upsilon$ results. Although we have shown comparisons for allowed transitions alone, it is clear that their results will be in even greater error for the forbidden transitions, for which the resonances are numerous, as shown in their Figs.~1 and 3. An indication of that is visible in their results for the 1--7, 1--8, and 1--39 transitions, listed in their Table~3.

\section{Conclusions}

By performing simple calculations in this short paper, we have demonstrated that the earlier reported $\Omega$ results of Mohan et al. \cite{mm1} are largely error free, but for $\Upsilon$ are overestimated by up to a factor of two, for several allowed transitions, mainly at lower temperatures. In addition, the behaviour of their $\Upsilon$ is not correct. Furthermore, they have reported $\Upsilon$ for less than 4\% of the transitions (among 52 levels) of Ba~XLVIII, whereas a complete set of data for all transitions is preferred in any diagnostic or modelling applications of plasmas. Similarly, in fusion plasmas for which such collisional data for this ion may be required, the prevailing temperatures are far higher, up to $\sim$10$^8$~K. Therefore, it is recommended that fresh calculations for this ion should be performed, for a much wider range of transitions and temperatures,  for the reliable adoption of collisional data for this ion. 

We will like to (re)emphasise again that the reliability of any calculation does not (much) depend on the (in)accuracy of the code adopted, but on its implementation. If a code is incorrectly and/or non-judicially applied then large discrepancies may occur, as demonstrated in this paper. A number of large discrepancies, for several atomic parameters,  have been noted earlier for many ions, and these have been  highlighted in our recent paper \cite{atom}, along with suggestions for their resolutions.


\end{document}